\documentclass[a4paper,fleqn,usenatbib]{mnras}
\usepackage{newtxtext,newtxmath}
\usepackage[T1]{fontenc}
\usepackage{ae,aecompl}

\usepackage{graphicx}	% Including figure files
\usepackage{amsmath}	% Advanced maths commands
\usepackage{amssymb}	% Extra maths symbols

\usepackage{subfigure}

\makeatletter
\let\jnl@style=\rm
\def\ref@jnl#1{{\jnl@style#1}}
\makeatother

\newcommand*\patchAmsMathEnvironmentForLineno[1]{%
\expandafter\let\csname old#1\expandafter\endcsname\csname #1\endcsname
\expandafter\let\csname oldend#1\expandafter\endcsname\csname end#1\endcsname
\renewenvironment{#1}%
{\linenomath\csname old#1\endcsname}%
{\csname oldend#1\endcsname\endlinenomath}}% 
\newcommand{\ngc}{{NGC\,5506 }}
\def\approxgt{\mathrel{\hbox{\rlap{\lower.55ex \hbox {$\sim$}}
        \kern-.3em \raise.4ex \hbox{$>$}}}}
\def\approxlt{\mathrel{\hbox{\rlap{\lower.55ex \hbox {$\sim$}}
        \kern-.3em \raise.4ex \hbox{$<$}}}}

\def\H0{{\rm ~km~s^{-1}~Mpc^{-1}}}

\def\p9{{Pa$9$}}

\def\L2-10{L$_{\rm 2-10keV}$}

\def\.25{0.25 keV\thinspace}

\def\d19{D$\,\leq\,$19~Mpc}

\newcommand{\nus}{\textit{NuSTAR}}
\newcommand{\nusc}{\nus\ }

\newcommand{\suz}{\textit{Suzaku}}
\newcommand{\suzc}{\suz\ }

\newcommand{\xmn}{\textit{XMM-Newton}}
\newcommand{\xmnc}{\xmn\ }

\newcommand{\cha}{\textit{Chandra}}
\newcommand{\chac}{\cha\ }

\newcommand{\fe}{Fe~K$\alpha$}

\usepackage{paralist}
\usepackage{fancyhdr,lastpage}
\usepackage{color,hyperref}
\usepackage{pslatex}
\usepackage{amssymb}
\usepackage{afterpage}
\usepackage{upgreek}

        {\begin{itemize}[#1]}{\end{itemize}%
         \vspace{-.6\baselineskip}}

        {\vspace{-\baselineskip}\begin{list}{#1}{%
        \setlength{\partopsep}{0pt}%
        \setlength{\topsep}{0pt}}}
        {\end{list}\vspace{-.6\baselineskip}}

        {\begin{compactitem}[#1]}{\end{compactitem}}

\title[NGC~5506 in X-ray]
{Multi-epoch analysis of the X-ray spectrum of the active galactic nucleus in NGC~5506}

\author[S. Sun et al]{Shangyu Sun$^{1,4}$, Matteo Guainazzi$^{2}$, Qingling Ni$^{1}$, Jingchun Wang$^1$, Chenyang Qian$^1$,\\ \newauthor Fangzheng Shi$^1$, Yu Wang$^1$, Cosimo Bambi$^{1,3}$\thanks{Corresponding author: bambi@fudan.edu.cn}\\
$^1$ Center for Field Theory and Particle Physics and Department of Physics, Fudan University, 2005 Songhu Road, Shanghai 200438, China \\
$^2$ ESTEC/ESA, Keplerlaan 1, 2201AZ Noordwijk, The Netherlands \\
$^3$ Theoretical Astrophysics, Eberhard-Karls Universit\"at T\"ubingen, Auf der Morgenstelle 10, 72076 T\"ubingen, Germany\\
$^4$ now at Shanghai Astronomical Observatory, 80 Nandan Road, Shanghai 200030, China\\
}

\pubyear{2016}
\pagerange{\pageref{firstpage}--\pageref{lastpage}}
\begin{document}
\maketitle
\label{firstpage}

%\date{Received / Accepted}

\begin{abstract} 

We present a multi-epoch X-ray spectroscopy analysis of the nearby narrow-line Seyfert\,I galaxy NGC\,5506. For the first time, spectra taken by \cha, \xmn, \suz, and \nusc - covering the 2000--2014 time span - are analyzed simultaneously, using state-of-the-art models to describe reprocessing of the primary continuum by optical thick matter in the AGN environment. The main goal of our study is determining the spin of the supermassive black hole (SMBH). The nuclear X-ray spectrum is photoelectrically absorbed by matter with column density $\simeq 3 \times 10^{22}$~cm$^{-2}$. A soft excess is present at energies lower than the photoelectric cut-off. Both photo-ionized and collisionally ionized components are required to fit it. This component is constant over the time-scales probed by our data. The spectrum at energies higher than 2 keV is variable. We propose that its evolution could be driven by flux-dependent changes in the geometry of the innermost regions of the accretion disk. The black hole spin in NGC\
,5506 is constrained to be 0.93$\pm  _{ 0.04 }^{0.04}$ at 90\% confidence level for one interesting parameter.
\end{abstract}

\begin{keywords}
Galaxies: active - Galaxies: Individual: NGC~5506 - Accretion, accretion discs
 \end{keywords}

\section{INTRODUCTION}\label{intro}

\ngc is a nearby ($z$=0.0062; $\sim 24$\,Mpc) edge-on Sa galaxy within the local Virgo supercluster, hosting an active galactic nucleus (AGN) classified as radio-quiet narrow-line Seyfert I \citep[NLSy1; ][]{2002A&A...391L..21N}. The NLSy1 galaxies generally have lower SMBH masses and higher accretion rates \citep{2007ApJ...667L..33K}. 
\ngc is an obscured AGN but still Compton-thin \citep[$N_H\approx 3\times 10^{22}{\rm cm}^{-2}$; ][]{1999ApJ...515..567W}.
Its brightness in X-ray makes it a proper source for studying in detail the high-energy activities near an active SMBH of this kind.

The hard X-ray ($E>2$~keV) emission from radio-quiet AGN (without any prominent jet) originates primarily from the vicinity of the central engine, the SMBH, as close as a few tens of gravitational radii, $r_{\rm g}$ ($r_{\rm g}=GM/c^{2}$; gravitational radius) away from the event horizon as suggested by the short variability timescales in hard X-ray band \citep[e.g.,][]{1992ApJS...82...93G,2006Natur.444..730M,2007ASPC..373..149U,2013ApJ...767..121Z}. 
In this innermost region of AGN, the accretion disc (AD), whose blackbody radiation spectrum peaks in the UV wavelengths, is illuminated by a hard X-ray source. The spectrum of this hard X-ray source is typically described by a power-law function with a high-energy cutoff. The hard X-ray source can be a magnetically dominated corona residing above the surface of the disc \citep[e.g.,][]{1979ApJ...229..318G,1991ApJ...380L..51H,1994ApJ...432L..95H,1998MNRAS.299L..15D,2001MNRAS.328..958M}, where the AD thermal photons are inverse-Compton up-scattered by relativistic electrons. Compton scattering, fluorescence and photoelectric absorption occur in the AD, creating the so called ''reflection spectrum''\citep{1991MNRAS.249..352G}. Iron fluorescence in a dense and relatively cold medium \citep[e.g.,][]{1985MNRAS.216P..65B,1989MNRAS.236P..39N,1990Natur.344..132P,1991MNRAS.249..352G,1994MNRAS.268..405N} produces an emission line at about 6.4\,keV. 
The shape pf the Fe K$_{\alpha}$ profile is affected by the kinematic Doppler shift due to the AD rotation, as well as by special and General relativistic effects due to the propagation of the photons in the curved space-time caused by the (possibly fast 
spinning) SMBH \citep{1989MNRAS.238..729F,1991ApJ...376...90L}. These relativistically broadened \fe\ lines in AGN spectra have been resolved by \cha, \xmn, \suz, and \nusc \citep[e.g.,][]{2003MNRAS.342..239B,2003PhR...377..389R,2007ARA&A..45..441M,2007MNRAS.382..194N,2009Natur.459..540F,2012MNRAS.423.3299N,2012MNRAS.422.2510W,2013MNRAS.428.2901W,2013Natur.494..449R,2014MNRAS.443.1723P}.
The spectral analysis of relativistically broadened lines can provide a measurement of the black hole spin \citep[e.g.,][]{2006ApJ...652.1028B,2008ApJ...675.1048R,2014MNRAS.443.1723P} and potentially even test Einstein's gravity~\citep{2017RvMP...89b5001B,2015ApJ...811..130J}.
%The black hole spin can be measured through the spectral analysis on these relativistically broadened lines \citep[e.g.,][]{2006ApJ...652.1028B,2008ApJ...675.1048R,2014MNRAS.443.1723P}.
% //  \\

In the soft X-ray band, an excess over the extrapolation of the absorbed nuclear emission has been observed in many obscured AGN spectra. The origin of this \emph{soft excess} is complex. The AGN nuclear emission can photoionize its circumnuclear gas, and can result in narrow radiative recombination continua (RRC) as well as recombination lines of highly-ionised atomic species. This kind of photoionized spectra have been found in several sources, e.g., the Circinus Galaxy \citep{2001ApJ...546L..13S}, Mrk~3 \citep{2000ApJ...543L.115S,2005MNRAS.360..380B,2005MNRAS.360.1123P}, NGC~1068 \citep{2002ApJ...575..732K,2001ApJ...556....6Y}, NGC~4151 \citep{2004MNRAS.350....1S}.
In a \xmnc survey of obscured AGNs \citep{2007MNRAS.374.1290G}, photoionisation signature was found in 36\% of the sample. 
The line widths of RRC show plasma temperatures of $\sim$eV \citep{2002ApJ...575..732K}.  In the spectra of some obscured AGNs, higher order transitions are enhanced with respect to pure photoionisation.
In these cases, resonant scattering is important in the balance between photoionisation and photoexcitation.  
Alternatively, shocks produced by AGN outflows \citep{2005ApJ...635L.121K} or starburst \citep{1998ApJ...501...94S,2001ApJ...546..845G} can also heat up ambient gas to $\sim 10^6$~K. A spectrum of collisionally ionized plasma is characteristic for these scenarios \citep{1989ApJ...339..689V}, but it contributes to observed soft X-ray emission only in galaxies with strong nuclear star formation \citep{2009A&A...505..589G}.

\cite{2003A&A...402..141B} analyzed \ngc X-ray spectra taken from 1997 to 2002, reporting only a narrow \fe\ line. However, deeper \xmnc observations unveiled a moderately broadened component of the \fe\ line \citep{2010MNRAS.406.2013G}. Furthermore, the spectrum observed by \nus, which has a high sensitivity above 10\,keV, constrained the high-energy cutoff of the continuum \citep{2015MNRAS.447.3029M}. In this paper, we analyze for the first time all the good quality X-ray archival data of NGC\,5506. The main goal of this paper is achieving the most accurate determination so far of the relativistic accretion disk and black hole parameters, taking advantage of the unprecedented combination of signal-to-noise ratio and energy broadband coverage (0.2--79\,keV).

In Sect.~\ref{obser}, we report the observations and data products from each instrument. In Sect.~\ref{analy}, we show the procedures and results on the spectral analysis. Then we discuss and conclude in Sect.~\ref{discuss}.

\section{OBSERVATIONS AND DATA PRODUCTS}\label{obser}
For the sake of this work, 11 deep observations (all those whose exposure time is larger than 20\,ks) in X-rays were selected from NASA's High Energy Astrophysics Science Archive Research Center (HEASARC). The observation time, id, exposure time, start time, and its corresponding epoch name used in this work (E1--E11) are listed in Table~\ref{Tobse}.

\begin{table*}
\caption{Log of the observations discussed in this paper.}\label{Tobse}
\centering
\begin{tabular}{lllll} 
\hline\hline            
Observatory  &   obsid       &    exposure &   start time          & epoch \\
             &               &    time$^{\clubsuit}$ [s]  &[yyyy-mm-dd hh:mm:ss] & index\\
\hline       
\cha         &   1598        &    90040     &   2000-12-31 06:21:12 &   E01   \\
\xmn         &   0013140101  &    20007     &   2001-02-02 22:01:45 &   E02   \\
\xmn         &   0201830201  &    21617     &   2004-07-11 09:47:01 &   E03   \\
\xmn         &   0201830301  &    20409     &   2004-07-14 22:11:39 &   E04   \\
\xmn         &   0201830401  &    21956     &   2004-07-22 13:07:37 &   E05   \\
\xmn         &   0201830501  &    20411     &   2004-08-07 20:17:40 &   E06   \\
\suz         &   701030010   &    47753     &   2006-08-08 16:30:05 &   E07   \\
\suz         &   701030020   &    53296     &   2006-08-11 02:26:18 &   E08   \\
\suz         &   701030030   &    57406     &   2007-01-31 02:12:12 &   E09   \\
\xmn         &   0554170101  &    88919     &   2008-07-27 07:20:37 &   E10   \\
\nus         &  60061323002  &    56585     &   2014-04-01 23:41:07 &   E11   \\
\hline    
\multicolumn{4}{l}{$^{\clubsuit}$before filtering and without considering dead time}\\
\end{tabular}
\end{table*}

\subsection{\xmn}\label{xmm}
The data from European Photon Imaging Camera (EPIC)- Metal Oxide Semi-conductor \citep[MOS;][]{2001A&A...365L..27T} and pn \citep{2001A&A...365L..18S} were reduced using the software \textsc{SAS} provided by ESA \citep{2004ASPC..314..759G}. Flaring particle background was filtered using light curves extracted at high energies (Pulse-Invariant (PI)$>$10 keV for MOS and PI between 10 keV and 12 keV for PN) and with single events only. We removed the high particle background intervals above the appropriate count rate thresholds (see Table \ref{DataRedTab}) 
individually set for each observation and each instrument \citep{2010MNRAS.406.2013G}.
We selected single, double, triple and quadruple events for MOS; single and double events for PN. Source spectra were extracted from circular regions around the source centroid. For MOS data, background spectra were extracted from annulus region with no extra sources around the same centroid as the corresponding source spectra. For PN, background spectra were extracted from circular regions at the same height in detector coordinates to ensure the same charge transfer inefficiency correction as the source spectra. The redistribution matrix and the effective area were computed through the \textsc{SAS} tasks, \textsc{arfgen} and \textsc{rmfgen}. The spectral energy range used in this work is 1--10\,keV.

Data from reflection grating spectrometer (RGS) were processed with \textsc{SAS} v.14.0.0, by using the standard task pipeline \textsc{rgsproc}. The spectral energy range used in this work is 0.2--2\,keV.

\begin{table}
\caption{Parameters for EPIC data reduction.}\label{DataRedTab}
\centering
\begin{tabular}{llll} 
\hline\hline            
ObsID   &  $C_{\rm th}$ $^{a}$& $ R_{\rm s}$ $^{b}$ \\
        &  [$\rm s^{-1}$]     & [arcsec]\\
\hline
0013140101    &  0.2 / 0.5    &   100 / 40\\
0201830201    &  0.5 / 0.5    &   42 / 45\\
0201830301    &  0.5 / 0.35   &   40 / 42\\
0201830401    &  0.35 / 0.35  &   40 / 42\\
0201830501    &  0.35 / 0.35  &   40 / 42\\
0554170101    &  0.35 / 0.35  &   50 / 50\\
\hline       
\multicolumn{3}{l}{$^{a}$ Threshold count rate on the high-energy, single-event}\\ \multicolumn{3}{l}{\quad  light curve to identify intervals of high particle}\\
\multicolumn{3}{l}{\quad background in MOS/PN, respectively.}\\
\multicolumn{3}{l}{$^{b}$ $R_{\rm s}$ Radius for the source spectrum extraction region.}\\
\end{tabular}
\end{table}

\subsection{\suz}\label{suz}

The Suzaku data were reduced by software package \textsc{HEASOFT} v6.16, using the \textsc{aepipeline } script following the \emph{Suzaku ABC Guide}\footnote{It can be found at \url{https://heasarc.gsfc.nasa.gov/docs/suzaku/analysis/abc/}.} to perform reprocessing (update energy calibration) and rescreening.

As for the X-ray imaging spectrometer (XIS) data, we extract spectra using a 260$\arcsec$ region around the source through \textsc{xselect}. The ancillary response matrices (ARFs) and the detector response matrices (RMFs) were generated by \textsc{xisrmfgen} and \textsc{xisarfgen} task pipelines, respectively. The task pipeline \textsc{addascaspec} was used to combine front-side illuminated spectra and responses. The spectra from XIS0, XIS2, and XIS3 are combined for each observation and referred to in the following as 'front-side illuminated spectra'. The spectrum from XIS1 is referred to in the following as 'back-side illuminated spectrum'. 
The spectral energy range used in this work is 0.5--10\,keV.

The data from the hard X-ray detector (HXD) include those from gadolinium silicate crystal counters (${\rm Gd}_2{\rm Si}{\rm O}_5$(Ce); GSO; $>$50\,keV) and from the silicon PIN diodes ($< 50$ \,keV). In this work, only PIN data are used, while \ngc was not detected by the GSO. The PIN data was reduced by standard task pipeline \textsc{hxdpinxbpi}. 
The spectral energy range used in this work is 10--50\,keV.

\subsection{\nus}\label{nus}
The data from focal-plane modules (FPM) 1 and 2 were reduced with \textsc{HEASOFT} v6.16, calibration files of \textsc{indx20150316}. The \textsc{nupipeline} and \textsc{nuproducts} tasks were run. Source spectra were extracted from a circular region with radius $2\arcmin$. An off-source background region of the same size was used. The two regions were separated by $1\arcmin$.
The spectral energy range used in this work is 5--79\,keV.

\subsection{\cha}\label{cha}
The \chac high energy grating (HEG) and medium energy grating (MEG) spectra in this work were extracted from the \emph{Chandra Grating-Data Archive and Catalog} (TGCat). Details on the data reduction can be found in \cite{2011AJ....141..129H}. The spectral energy ranges used in this work are 0.4--5\,keV for the two medium energy grating (MEG) spectra, and 0.8--10\,keV for the two high energy grating (HEG) spectra.

\subsection{Spectral binning}\label{bin}
%Spectra were re-binned to ensure that each background-subtracted energy channel had at least one count, and to sample the instrumental energy resolution according to the prescription in \cite{2016A&A...587A.151K}.

We applied the same spectral binning criterion to all spectra, CCD- and high-resolution alike. We choose the statistically-motivated recipe in \cite{2016A&A...587A.151K} that provides a criterion dependent solely on the number of counts per resolution element, and can therefore be applied to all types of X-ray spectra. Additional rebinning was applied only post spectral analysis to generate figures, whose residuals are clearly distinguishable in each plot.

\section{SPECTRAL RESULTS}\label{analy}

The spectral analysis was performed using a two-tier strategy. Firstly, we fit the high-resolution spectra only in the $E<$1.3 keV energy range, dominated by the epoch-invariant soft excess (Sect.~\ref{RGStest}). Then, we fit the broadband spectrum, including in the spectral model the components fitting the soft excess with all their parameters frozen to their best fit-values (Sect.~\ref{model}). The spectral fits are performed using XSPEC v.12.8.2 \citep{1996ASPC..101...17A}. The statistical uncertainties reported in this paper corresponds to 90\% confidence level for one interesting parameter \citep{1976ApJ...208..177L}.

\subsection{Spectral evolution across 11 epochs}\label{}
\begin{figure*}
\centering
\includegraphics[width=.70\textwidth,angle=270]{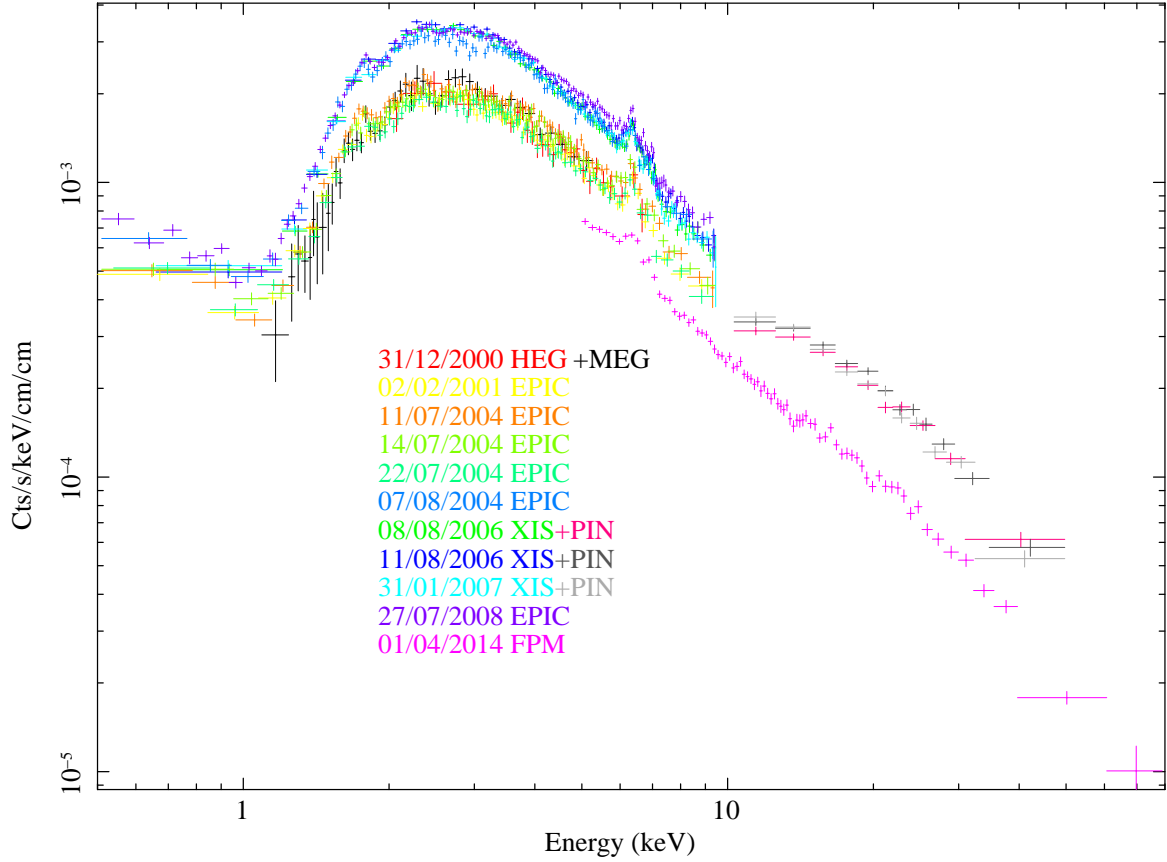}
\caption{\ngc count spectra from  \cha, \xmn, \suz, and \nusc (corrected by the effective areas). The observation dates, instruments, and corresponding colors are labeled in the figure. 
 }
\label{Fvar}
\end{figure*}

Fig.~\ref{Fvar} illustrates the X-ray spectral variability in \ngc from epoch~1 to 11. Variability is observed only above 1.3\,keV. The spectral variation is a factor of 4 at 5\,keV between the highest and lowest state. These multi-epoch spectra can be classified into three states according to the flux level: high states (A) including E06--E10, intermediate states (B): E01--E05, low state (C): E11. \ngc evolved from intermediate states (E01--E05), to high states (E06--E10), and reached finally a low state (E11).

%For each epoch, each energy coverage, only one spectrum is shown. E01:\,MEG1(green;0.4--5\,keV), HEG1(light blue;0.8--10\,keV); E02:\,EPIC1a(purple;0.3--10\,keV); E03:\,EPIC1b(deep grey;0.3--10\,keV); E04:\,EPIC1c(black;0.3--10\,keV); E05:\,EPIC1d(green;0.3--10\,keV); E06:\,EPIC1e(light blue;0.3--10\,keV); E07:\,XIS1a(yellow;0.5--10\,keV); E08:\,XIS1b(orange;0.5--10\,keV); E09:\,XIS1c(light green;0.5--10\,keV); E10:\,EPIC1f(yellow;0.3--10\,keV); E11:\,FPM1(Turkish blue;5--79\,keV); RGS1(black;0.2--2\,keV), RGS2(red;0.2--2\,keV)

\subsection{The soft X-ray spectrum}\label{RGStest}

%In order to achieve a broad-band modeling of the NGC5506 X-ray spectra, we considered initially the soft X-ray band ($E <$1.3\,keV) only. 
Our working hypothesis is that soft X-rays are dominated by emission due to diffuse gas associated to extended Narrow Line Regions, while emission at higher energy is dominated by the nuclear emission. Different models of the soft X-ray spectrum are compared in Tab.~\ref{T:soft}. Comparing models of increasing level of complexity via the F-test, we define the best-fit baseline model of the soft X-ray spectrum as: {\rm \textsc{powerlaw}}[1] + {\rm \textsc{xstar2xspec}} + {\rm \textsc{apec}}[1]  + {\rm \textsc{apec}}[2] to model the soft X-ray component.     
Each component is described in the following. 
%\begin{description}

%\item $\bullet$ 
\textsc{xstar2xspec} is an additive table model generated by \textsc{XSTAR}. \textsc{XSTAR} \citep{2001ApJS..133..221K} calculates the physical conditions and emission spectra of photoionized gases. It is parameterized by a column density $N_{\rm H}^{\rm phot.}$ and ionisation $(\xi_{\rm phot.})$. \textsc{powerlaw} component (with photon index $\alpha_{\rm phot.}$) is used for approximating free electron scattering in the photoionized gas. \textsc{apec} \citep{2001ApJ...556L..91S} components are emission spectra from collisionally-ionized diffuse gases of temperatures $kT_1^{\rm coll.}$ and $kT_2^{\rm coll.}$ calculated using the ATOMDB code v2.0.2.\footnote{\url{http://atomdb.org/}}. 
%\end{description}

\begin{table}
\small{
\caption{Models for high-resolution spectra below 1.3\,keV.}\label{T:soft}
\centering
\begin{tabular}{lrrc} 
\hline\hline
model & C-Statistic & $\chi^2 /n.d.f.$ & F-test prob.$^{\clubsuit}$\\                     
\hline
A({\scriptsize nucl.}) & 18593 & 16847/5496=3.065 & --\\  
B({\scriptsize nucl. + coll.}) &  6771 &  6412/5493=1.167           & $4.0\times 10^{-17}$[A]\\
C({\scriptsize nucl. + phot.}) &  6356 &  6185/5491=1.126           & $9.5\times 10^{-11}$[A]\\
D({\scriptsize nucl. + phot. + coll.}) &  6186 &  6071/5488=1.106   & $3.6\times 10^{-8}$ [C]\\
E({\scriptsize nucl. + phot. + 2 coll.}) &  6160 &  6030/5485=1.099 & $2.2\times 10^{-7}$ [D]\\ 
\hline 
\multicolumn{4}{l}{A: {\scriptsize {\rm \textsc{tbabs}}[1] * ( {\rm \textsc{tbabs}}[2]*{\rm \textsc{powerlaw}}[2])}}\\
\multicolumn{4}{l}{B: {\scriptsize {\rm \textsc{tbabs}}[1] * ( {\rm \textsc{apec}} + {\rm \textsc{tbabs}}[2]*{\rm \textsc{powerlaw}}[2])}}\\
\multicolumn{4}{l}{C: {\scriptsize {\rm \textsc{tbabs}}[1] * ({\rm \textsc{powerlaw}}[1] + {\rm \textsc{xstar2xspec}} + {\rm \textsc{tbabs}}[2]*{\rm \textsc{powerlaw}}[2])}}\\
\multicolumn{4}{l}{D: {\scriptsize {\rm \textsc{tbabs}}[1] * ({\rm \textsc{powerlaw}}[1] + {\rm \textsc{xstar2xspec}} + {\rm \textsc{apec}} + {\rm \textsc{tbabs}}[2]*{\rm \textsc{powerlaw}}[2])}}\\
\multicolumn{4}{l}{E: {\scriptsize {\rm \textsc{tbabs}}[1] * ({\rm \textsc{powerlaw}}[1] + {\rm \textsc{xstar2xspec}} + {\rm \textsc{apec}}[1] + {\rm \textsc{apec}}[2] + {\rm \textsc{tbabs}}[2]*{\rm \textsc{powerlaw}}[2])}}\\
\multicolumn{4}{l}{See description for each model in Subsect.~\ref{RGStest}.}\\
\multicolumn{4}{l}{number of PHA bins: 5500}\\
\multicolumn{4}{l}{$^{\clubsuit}$ comparing with the null model in [ ]}\\
%\multicolumn{4}{l}{\quad  the null model in [ ]}\\
\end{tabular}                                           
}
\end{table}

% 
% 
%                                                           C-Statistic  degrees of freedom Chi-Squared   Reduced chi-squared  probability number of parameters compare with   
% E TBabs*(TBabs*powerlaw +powerlaw + NGC5506 +apec +apec)    6160        5485                6030         1.099                2.217251e-07  15                D                  
% D TBabs*(TBabs*powerlaw +powerlaw + NGC5506 +apec)          6186        5488                6071         1.106                3.552035e-08  12                C
% ##################################################################################
% C TBabs*(TBabs*powerlaw +powerlaw + NGC5506)                6356        5491                6185         1.126                9.498405e-11   9                A
% B TBabs*(TBabs*powerlaw +apec)                              6771        5493                6412         1.167                4.026067e-17   7                A
% ##################################################################################
% A # model  TBabs*( TBabs*powerlaw)                         18593        5496               16847         3.0653               0.000000e+00   4                -
% 5500 PHA bins

To test if the soft X-ray emission is constant during our observations (as expected on astrophysical grounds; see Sect.~\ref{intro}), the RGS spectra of each epoch were fit using the aforementioned model. Constant-functions were fit to the best-fit soft X-ray parameters in each epoch as a function of time, and the probability of this assumption was calculated. The results of this test are reported in Tab.~\ref{T:RGS}. The high $p$-values of fits suggest that epoch-invariant model parameters describe the soft RGS spectra properly. To reduce the statistical uncertainties in RGS measurements, spectra from all the epochs were merged through the \textsc{SAS} task. Hereafter, the parameters of the model describing the soft X-ray band are assumed to be epoch-independent. The best-fit parameters of the photoionized and collisionally-ionized gases are reported in Tab.~\ref{Tfit}.

\begin{table}
\small{
\caption{Constant-function fitting to RGS epoch-dependent parameters}\label{T:RGS}
\centering
\begin{tabular}{lll} 
\hline\hline
parameter & $\chi^2/n.d.f.$ & p-value \\                     
\hline
$\alpha_{\rm phot.}$       &  4.07/5    &   0.539 \\   %&  0.393/5     &  0.9955
$kT_1^{\rm coll.}$         &  3.64/5    &   0.603 \\   %&  2.14/5      &  0.8300
$Z$                        &  0.65/5    &   0.986 \\   %&  1.31/5      &  0.9343
$kT_2^{\rm coll.}$         &  0.50/5    &   0.992 \\   %&  0.499/5     &  0.9921
log$(\xi_{\rm phot.})$     &  4.97/5    &   0.420 \\   %&  13.3/5      &  0.0207
\hline 
%\multicolumn{5}{l}{$^{\clubsuit}$} $^{\spadesuit}$\\
\multicolumn{3}{l}{number of entries: 6 (epochs)}\\
\multicolumn{3}{l}{number of parameter: 1 (constant-function)}\\
\multicolumn{3}{l}{See description for each parameter in Subsect.~\ref{model}.}\\
\end{tabular}                                           
}
\end{table}

\subsection{Broadband X-ray modeling}\label{model}

We then fit the X-ray broadband spectra of the 11 epochs using the following model:
% \begin{equation*}\label{fmodel}
% \normalsize{
%  \begin{array}{ll}
% {\rm \textsc{const}}[1] * {\rm \textsc{tbabs}}[1] & \\
% * ({\rm \textsc{powerlaw}}  + {\rm \textsc{apec}}[1] + {\rm \textsc{apec}}[2] +  {\rm \textsc{xstar2xspec}}   & \\
% + {\rm \textsc{tbabs}}[2]*[ {\rm \textsc{relxill}} * {\rm \textsc{MT}_{abs}}  + {\rm \textsc{const}}[2]*{\rm \textsc{MT}_{sca}}&  \\
% + {\rm \textsc{const}}[3]*{\rm \textsc{gsmooth}}*{\rm \textsc{MT}_{lin}} ] )&.\\
%   \end{array}
% }  
% \end{equation*}
\begin{equation*}\label{fmodel}
%\normalsize{
\small{
 \begin{array}{ll}
{\rm \textsc{const}} * {\rm \textsc{tbabs}}[1] * ({\rm \textsc{powerlaw}}  + {\rm \textsc{apec}}[1] + {\rm \textsc{apec}}[2]  +  {\rm \textsc{xstar2xspec}} & \\
  + {\rm \textsc{tbabs}}[2]*[ {\rm \textsc{relxill}} + {\rm \textsc{MT}_{sca}} + {\rm \textsc{MT}_{lin}} ] ) &.  \\
%+ {\rm \textsc{gsmooth}}*{\rm \textsc{MT}_{lin}} ] )&.\\
  \end{array}
}  
\end{equation*}
\textsc{xstar2xspec}, \textsc{powerlaw}, and \textsc{apec}, are explained in Subsect.~\ref{RGStest}. The parameters in these components are frozen at the values derived in Subsect.~\ref{RGStest} except the parameter elemental abundance ($Z$; see the description below in this subsection). The other components are described in the following. 
\begin{description}

\item $\bullet$ \textsc{Mytorus}\footnote{\url{http://mytorus.com/}} \citep{2009MNRAS.397.1549M} was used to calculate the reflection and scattering from optically thick matter in a toroidal geometry surrounding the AGN. In this work, this torus model set includes \textsc{MT}$_{\rm sca}$ and \textsc{MT}$_{\rm lin}$. \textsc{MT}$_{\rm sca}$ calculates the scattered continuum, i.e. the escaping continuum photons that have been scattered in the medium at least once. It is also referred to as the reflection spectrum. \textsc{MT}$_{\rm lin}$ calculates the fluorescent emission-line spectrum, which is produced by the zeroth-order (or unscattered) component resulting from the absorption of a continuum photon above the K-edge threshold as well as by the emission-line photons that are scattered before they escape the medium. It includes the Compton shoulder of the observed emission line. The column density of the reprocessing matter $N_{\rm H}^{\rm torus}$, the inclination angle $\theta_{\rm torus}$, and the 
photon index of the reprocessed primary continuum $\alpha_{\rm torus}$, are assumed epoch-invariant, following \cite{2010MNRAS.406.2013G}. The normalisation parameters of \textsc{MT}$_{\rm sca}$ and \textsc{MT}$_{\rm lin}$ ($N_{\rm sca}^{\rm torus}$ and $N_{\rm lin}^{\rm torus}$) are assumed to be independent free parameters in the fit (not linked to the primary nuclear flux), and epoch-invariant because the torus reprocessing is distant from the nuclear emission.

\item $\bullet$ \textsc{relxill} \citep{2014MNRAS.444L.100D,2014ApJ...782...76G} describes an X-ray source of a power-law photon index $\alpha_{\rm coro.}$, a normalisation factor $N_{\rm prim.}^{\rm coro.}$, and a cutoff at high energies $E_{\rm cut}^{\rm coro.}$ plus the X-ray reflection from a relativistically rotating disk in a strong gravitational field generated by a (possibly spinning) black hole. We assumed that the emission occurs in an annulus over the accretion disk, whose outer radius was fixed to 400\,$r_{\rm g}$, and whose inner radius coincides with the innermost stable circular orbit (ISCO), which is a function of the BH spin $a$ \citep{1972ApJ...178..347B}. We assumed that the radial profile of the emissivity function can be described by a single power-law of index $\beta_{\rm AD}$. The inclination angle of the AD is $\theta_{\rm AD}$. The definition of parameter \emph{reflection fraction} $F_{\rm refl.}$ can be found in \citep{2014MNRAS.444L.100D}. The following astrophysical parameters: $\alpha_{\rm coro.}$, $\beta_{\rm AD}$, $F_{\rm refl.}$, ionisation parameter of AD $\xi_{\rm AD}$, and $N_{\rm prim.}^{\rm coro.}$ are assumed independently variable in each epoch, while the other parameters are left epoch-invariant.

\item $\bullet$ \textsc{const} represents the cross-instrument calibration factors. It is left free to vary independently at each epoch.

\item $\bullet$ \textsc{tbabs}[1] describes obscuration due to intervening gas in our Galaxy by a hydrogen column density $N_{\rm H}^{\rm gala.}$. It is 
left epoch-invariant.

\item $\bullet$ \textsc{tbabs}[2] describes the obscuration of the nuclear emission by a hydrogen column density $N_{\rm H}^{\rm nucl.}$. It is 
left free to vary independently at each epoch.

\item $\bullet$ We assumed a variable and epoch-invariant elemental abundance $Z$. Its value is assumed to be the same across models (\textsc{apec}[1], \textsc{apec}[2], and \textsc{relxill}).

\end{description}

\subsubsection{Fit results}\label{}

\begin{figure*}
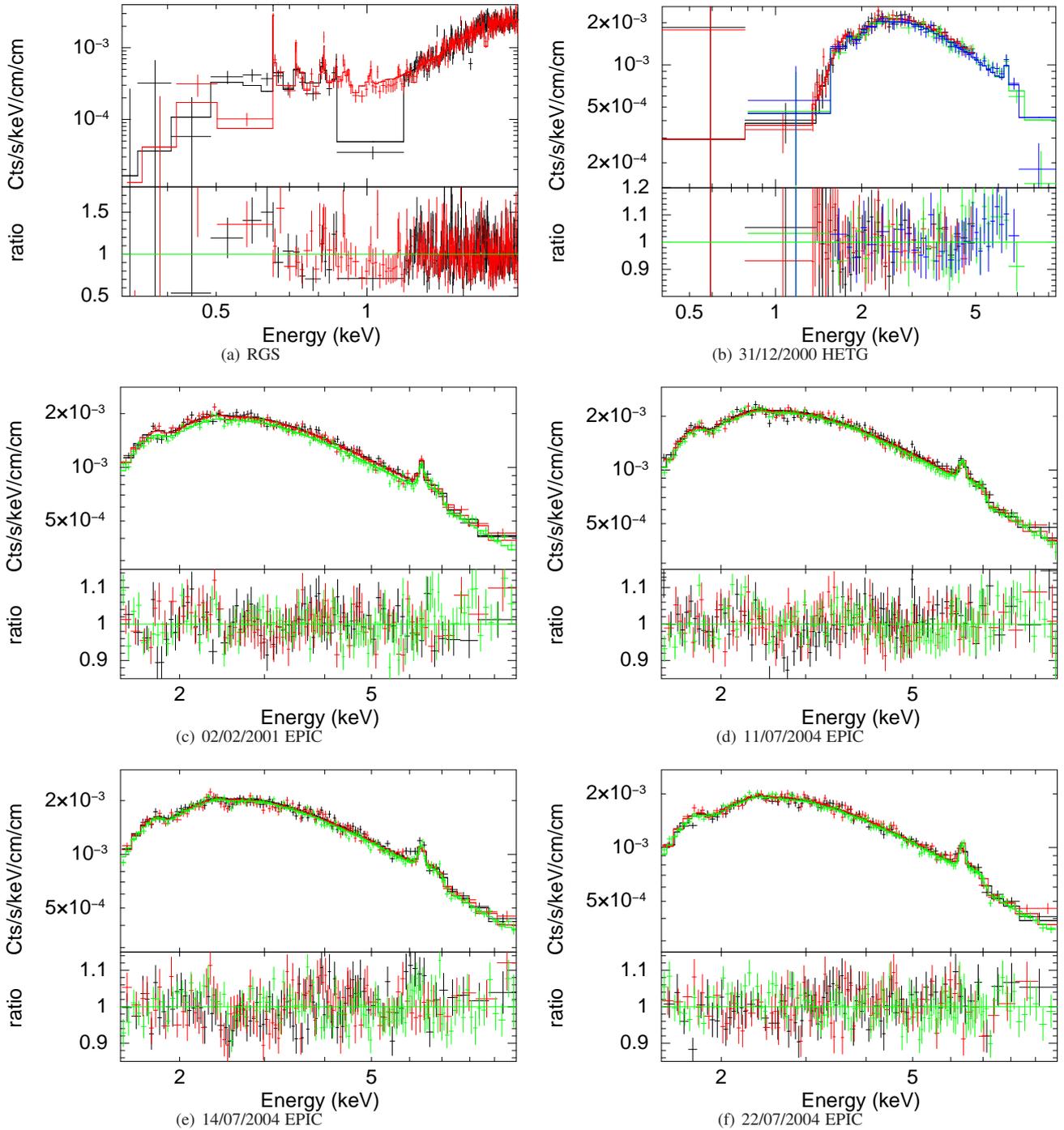

  \centering
  \subfigure[RGS]{
    \includegraphics[width=5.6cm,angle=270]{./I5plotH2-32_LdataRatio_RGS_Sig5-500.eps}}
  \subfigure[31/12/2000 HETG]{
    \includegraphics[width=5.6cm,angle=270]{./I5plotH2-32_LdataRatio_HETG_Sig3_10-100.eps}}  
    
  \subfigure[02/02/2001 EPIC]{
    \includegraphics[width=5.6cm,angle=270]{./I5plotH2-32_LdataRatio_EPICa.eps}}
  \subfigure[11/07/2004 EPIC]{
    \includegraphics[width=5.6cm,angle=270]{./I5plotH2-32_LdataRatio_EPICb.eps}}

  \subfigure[14/07/2004 EPIC]{
    \includegraphics[width=5.6cm,angle=270]{./I5plotH2-32_LdataRatio_EPICc.eps}}
  \subfigure[22/07/2004 EPIC]{
    \includegraphics[width=5.6cm,angle=270]{./I5plotH2-32_LdataRatio_EPICd.eps}}    
    
  \caption{\emph{Upper panels:} \ngc X-ray spectra from RGS and Epochs 1--5, and their best-fit models. The instrument name and observation date for each epoch are on the top the panel. Crosses: data; histogram lines: models. The data have been rebinned for visual clarity.} \emph{Lower panels:} the data/model ratios.
   \label{Fe00}
\end{figure*}

\begin{figure*}
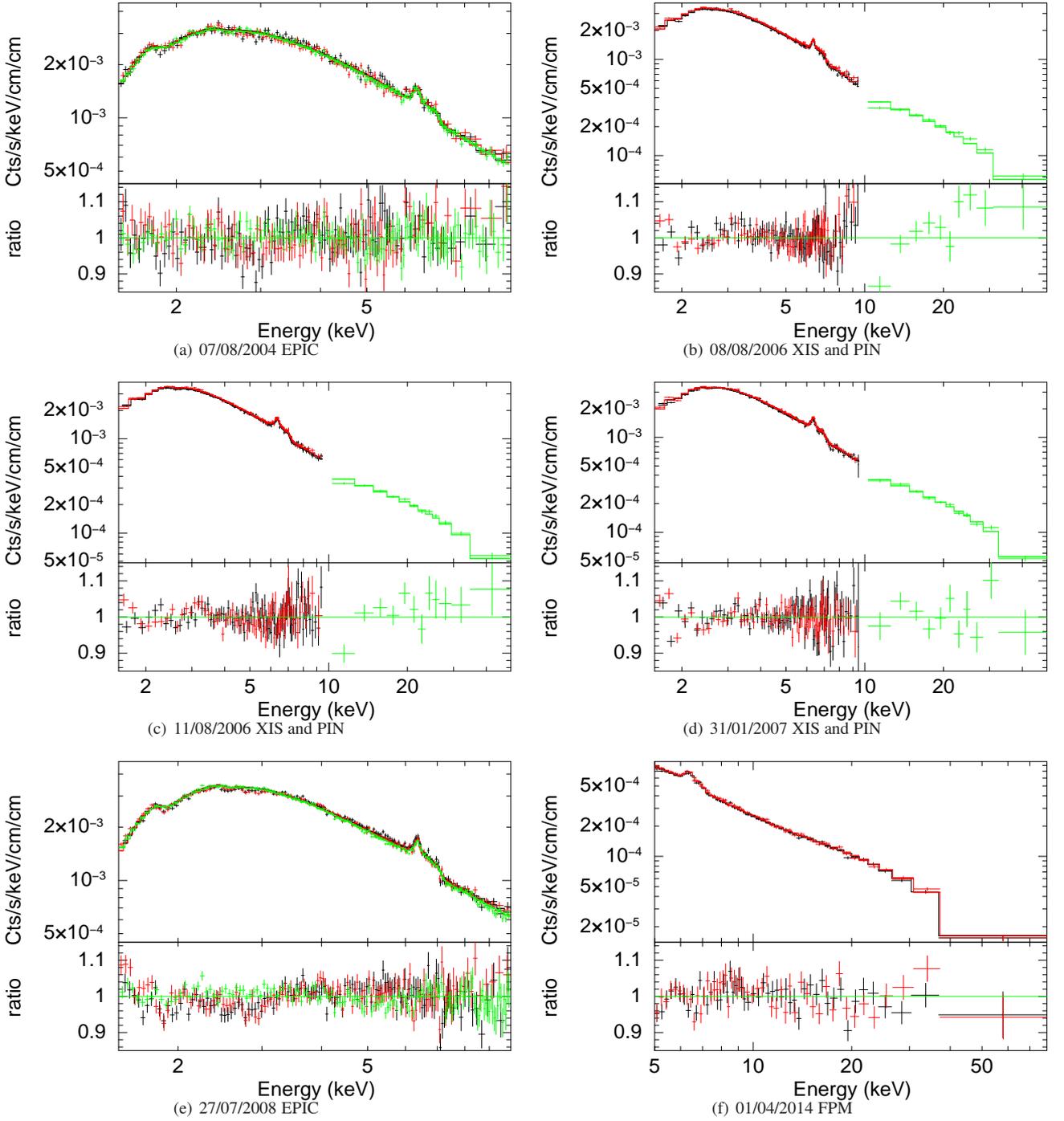

  \centering
  \subfigure[07/08/2004 EPIC]{
      %\label{fig:}
    \includegraphics[width=5.6cm,angle=270]{./I5plotH2-32_LdataRatio_EPICe.eps}}
  \subfigure[08/08/2006 XIS and PIN]{
    \includegraphics[width=5.6cm,angle=270]{./I5plotH2-32_LdataRatio_XISaPINa.eps}}  
    
  \subfigure[11/08/2006 XIS and PIN]{
    \includegraphics[width=5.6cm,angle=270]{./I5plotH2-32_LdataRatio_XISbPINb.eps}}
  \subfigure[31/01/2007 XIS and PIN]{
    \includegraphics[width=5.6cm,angle=270]{./I5plotH2-32_LdataRatio_XIScPINc.eps}}

  \subfigure[27/07/2008 EPIC]{
    \includegraphics[width=5.6cm,angle=270]{./I5plotH2-32_LdataRatio_EPICf.eps}}
  \subfigure[01/04/2014 FPM]{
    \includegraphics[width=5.6cm,angle=270]{./I5plotH2-32_LdataRatio_FPM_Sig30.eps}}    
    
  \caption{\ngc X-ray spectra from Epochs 6--11, and their best-fit models. See description in the caption of Fig.\ref{Fe00}.}
   \label{Fe11}
\end{figure*}

Spectra are shown epoch by epoch in the upper panels of Figs.~\ref{Fe00} to \ref{Fe11}, and their data/model ratios in the lower panels.

The the best-fit results of stable (epoch-invariant) astrophysical properties are listed in Tab.~\ref{Tfit}. The parameters that vary with epochs are in Tab.~\ref{Tevo2}.

% 
% \begin{figure*}
% \centering
% \includegraphics[width=.99\textwidth]{./F1.eps}
% \caption{\emph{Upper panels:} \ngc X-ray spectra from RGS and Epochs 1--5, and their best-fit models. The instrument name and observation date for each epoch are on the top the panel. Crosses: data; histogram lines: models. \emph{Lower panels:} the data/model ratios.}
% \label{Fe00}
% \end{figure*}
% 
% \begin{figure*}
% \centering
% \includegraphics[width=.99\textwidth]{./F2c.eps}
% \caption{\ngc X-ray spectra from Epochs 6--11, and their best-fit models. See description in the caption of Fig.\ref{Fe00}.
% }
% \label{Fe11}
% \end{figure*}

\begin{table}
\caption{Fit results of epoch-independent parameters.}\label{Tfit}
\centering
\small{
\begin{tabular}{lllll} \hline\hline 
 Par.  &  Par.   &  best-fit                &  \\                                                           %    Par.& 
 name  &  unit   &  value                   &    \\ \hline                                                              %    no. & 
 $N_{\rm H}^{\rm gala.}$ &${\rm cm}^{-2}$              &   3.5$\pm  _{  0.5 }^{0.6  }$  &  $\times 10^{21}$ \\             %      2 & 
& & & \\                                                  
$\alpha_{\rm phot.}$ &                             &   1.7$\pm  _{0.4 }^{0.3  } $ &                    \\                   %      3 & 
& & & \\ 
$kT_1^{\rm coll.}$    &  keV                        &  1.50$\pm  _{0.05}^{0.08  } $ &  $\times 10^{-1}$  \\                    %      7 & 
& & & \\ 
$Z$ & solar abundance&    0.50$\pm  _{  0.11 }^{0.04  }$  &             \\                     %      8 & 
& & & \\                    
$kT_2^{\rm coll.}$  &  keV                        &   4.1$\pm  _{0.3}^{0.4  } $ &  $\times 10^{-1}$ \\                  %     11 & 
& & & \\ 
% $N_{\rm H}^{\rm phot.}$  &                             & 2.0 $\pm  _{0.6  }^{1.1  }$    & $\times 10^{20}$   \\              %        & 
% & & & \\ 
log$(\xi_{\rm phot.})$  &                             &  1.5$\pm  _{0.2}^{0.2  }$  &                 \\                    %        & 
& & & \\ 
$a$    &                             &   0.93$\pm  _{ 0.04 }^{0.04}$  &         \\                       %     23 & 
& & & \\                                  
$\theta_{\rm AD}$    &  deg                        &   42 $\pm  _{3     }^{1    }$ &                           \\          %     24 & 
& & & \\                                                 
$E_{\rm cut}^{\rm coro.}$   &  keV                        &   5.00 $\pm  _{ 0.24  }^{0.10  }$  &  $\times 10^{2}$   \\           %     31 & 
& & & \\                                                     
$N_{\rm H}^{\rm torus}$ &${\rm cm}^{-2}$              &   1.04 $\pm  _{ 0.16  }^{0.06  }$   &  $\times 10^{25}$  \\              %     34 & 
& & & \\                                                  
$\theta_{\rm torus}$  &  deg                        &   42.5$\pm  _{ 2.5 }^{1.1 }$ &                         \\             %     35 & 
& & & \\                                                
$\alpha_{\rm torus}$ &                             &   1.97 $\pm  _{  0.05 }^{ 0.03 }$ &                    \\                   %     40 & 
& & & \\                                               
$N_{\rm sca}^{\rm torus}$   &   photons keV$^{-1}$cm$^{-2}$s$^{-1}$   &   5.19$\pm  _{ 1.75  }^{1.79  }$ &  $\times 10^{-2}$  \\          %  
& & & \\ %     54 & 
$N_{\rm lin}^{\rm torus}$   &  photons keV$^{-1}$cm$^{-2}$s$^{-1}$     &   4.55$\pm  _{ 1.35  }^{1.31  }$ &  $\times 10^{-5}$  \\          %     
& & & \\
$E_1^{\rm line}$   &  keV                        &   1.75$\pm  _{ 0.02  }^{0.01  }$    &                    \\
& & & \\ 
$\sigma_1^{\rm line}$   &  keV                        &   1.0$\pm  _{ 0.1  }^{0.1  }$ &  $\times 10^{-1}$  \\          %     50 & 
& & & \\                                                 
$E_2^{\rm line}$   &  keV                        &   6.89$\pm  _{ 0.01 }^{0.01 }$ &                    \\                              %     53 &  
& & & \\                                              
$\sigma_2^{\rm line}$ &  keV                        &   1.9$\pm  _{ 0.1  }^{0.1  }$ &  $\times 10^{-1}$  \\ 
& & & \\                                                
\hline 
\multicolumn{4}{l}{$^{\bigstar}$ See description for each parameter in Subsect.~\ref{model}.}\\
%\multicolumn{4}{l}{$^{\clubsuit}$ Total photons cm$^{-2}$s$^{-1}$ in the line.}\\
\end{tabular}  
}
\end{table}

\begin{table*}
\caption{Best-fit values of the time-dependent parameters.}\label{Tevo2}
\centering
\begin{tabular}{ccccccccc} 
\hline\hline             
epoch & start time        & energy flux (2--10 keV)$^{\spadesuit}$ &   $N_{\rm H}^{\rm nucl.}$       &  $\beta_{\rm AD}$ &  $\alpha_{\rm coro.}$  &  log$(\xi_{\rm AD})$   & $F_{\rm refl.}$   &  $F_{\rm prim.}^{\rm coro.}$ $^{\clubsuit}$\\   
      & [MJD]      & [$10^{-11} {\rm erg} {\rm s}^{-1} {\rm cm}^{-2}$]   &[$10^{22} {\rm cm}^{-2}$]  &  &          & &   &  [${\rm ph.} {\rm s}^{-1} {\rm cm}^{-2}$] \\   
\hline                                                                                                                            
E01  & 51909 & 6.17$\pm _{0.04}^{0.04}$    &   3.06  $\pm  _{ 0.06 }^{ 0.05}$  &  2.50 $\pm  _{0.10  }^{0.08  }$  &  2.05 $\pm  _{0.06}^{  0.04}$  &  3.16 $\pm  _{0.04}^{0.01}$  &   1.35 $\pm _{0.05}^{0.14}$  &  7.51 $\pm _{0.08}^{0.08} \times 10^{-1}$  \\ 
                  &       &  &       &  &          & &   &       \\ 
E02  & 51942 & 6.28$\pm _{0.03}^{0.03}$    &   3.03  $\pm  _{0.22  }^{0.24 }$  &  3.00 $\pm  _{0.10  }^{0.08  }$  &  1.88 $\pm  _{0.05 }^{0.11 }$  &  2.70 $\pm  _{0.18}^{0.14}$  &   1.18 $\pm _{0.12}^{0.18}$  &  1.11 $\pm _{0.01}^{0.01}$  \\
                  &       &  &       &  &          & &   &       \\ 
E03  & 53197 & 6.99$\pm _{0.03}^{0.03}$    &   2.87  $\pm  _{0.05  }^{ 0.10}$  &  3.12 $\pm  _{0.11  }^{0.09  }$  &  1.73 $\pm  _{ 0.05}^{ 0.07}$  &  3.13 $\pm  _{0.04}^{0.01}$  &   1.50 $\pm _{0.02}^{0.02}$  &  1.65 $\pm _{0.01}^{0.01}$  \\ 
                  &       &  &       &  &          & &   &       \\ 
E04  & 53200 & 6.75$\pm _{0.03}^{0.03}$    &   2.90  $\pm  _{0.07  }^{0.03 }$  &  3.12 $\pm  _{0.10  }^{0.08  }$  &  1.85 $\pm  _{0.06 }^{0.04 }$  &  3.00 $\pm  _{0.17}^{0.04}$  &   1.50 $\pm _{0.08}^{0.03}$  &  1.60 $\pm _{0.01}^{0.01}$  \\ 
                  &       &  &       &  &          & &   &       \\ 
E05  & 53208 & 6.05$\pm _{0.03}^{0.03}$    &   2.96  $\pm  _{0.12  }^{0.07 }$  &  2.04 $\pm  _{0.25  }^{0.21  }$  &  1.85 $\pm  _{0.05 }^{0.03 }$  &  2.75 $\pm  _{0.11}^{0.16}$  &   1.50 $\pm _{0.05}^{0.02}$  &  1.81 $\pm _{0.01}^{0.01}$  \\ 
                  &       &  &       &  &          & &   &       \\ 
E06  & 53234 & 9.83$\pm _{0.04}^{0.04}$    &   2.79  $\pm  _{0.03  }^{0.03 }$  &  3.12 $\pm  _{0.08  }^{0.06  }$  &  1.78 $\pm  _{ 0.06}^{ 0.05}$  &  3.50 $\pm  _{0.15}^{0.04}$  &   1.48 $\pm _{0.02}^{0.03}$  &  2.54 $\pm _{0.02}^{0.02}$  \\
                  &       &  &       &  &          & &   &       \\ 
E07  & 53955 & 9.81$\pm _{0.03}^{0.03}$    &   3.30  $\pm  _{0.20  }^{0.20 }$  &  1.38 $\pm  _{0.06  }^{0.04  }$  &  2.25 $\pm  _{0.11}^{ 0.13 }$  &  3.20 $\pm  _{0.12}^{0.10}$  &   1.50 $\pm _{0.12}^{0.04}$  &  2.16 $\pm _{0.01}^{0.01}$  \\ 
                  &       &  &       &  &          & &   &       \\ 
E08  & 53958 &10.37$\pm _{0.03}^{0.03}$    &   3.22  $\pm  _{0.14 }^{ 0.15 }$  &  1.38 $\pm  _{0.15  }^{0.11  }$  &  2.06 $\pm  _{0.11}^{  0.07}$  &  3.35 $\pm  _{0.06}^{0.08}$  &   1.18 $\pm _{0.10}^{0.18}$  &  2.24 $\pm _{0.01}^{0.02}$  \\ 
                  &       &  &       &  &          & &   &       \\ 
E09  & 54131 & 9.87$\pm _{0.03}^{0.03}$    &   3.10  $\pm  _{0.06 }^{ 0.01 }$  &  2.24 $\pm  _{0.13  }^{0.11  }$  &  2.14 $\pm  _{0.08}^{ 0.08 }$  &  3.26 $\pm  _{0.09}^{0.04}$  &   1.50 $\pm _{0.12}^{0.04}$  &  1.93 $\pm _{0.01}^{0.01}$  \\ 
                  &       &  &       &  &          & &   &       \\ 
E10  & 54833 &10.89$\pm _{0.02}^{0.02}$    &   2.91  $\pm  _{0.04  }^{0.07 }$  &  2.24 $\pm  _{0.03  }^{0.10  }$  &  2.21 $\pm  _{ 0.07}^{0.11 }$  &  3.50 $\pm  _{0.04}^{0.01}$  &   1.35 $\pm _{0.04}^{0.04}$  &  2.57 $\pm _{0.02}^{0.01}$  \\ 
                  &       &  &       &  &          & &   &       \\ 
E11  & 56748 & 5.83$\pm _{0.02}^{0.02}$    &   3.25  $\pm  _{1.0   }^{1.1  }$  &  1.40 $\pm  _{0.64  }^{0.75  }$  &  2.02 $\pm  _{ 0.10}^{ 0.08}$  &  2.69 $\pm  _{0.07}^{0.03}$  &   1.13 $\pm _{0.03}^{0.05}$  &  7.44 $\pm _{0.04}^{0.04} \times 10^{-2}$  \\                  
\hline
\multicolumn{9}{l}{See description for each parameter in Subsect.~\ref{model}.}\\
\multicolumn{9}{l}{$^{\spadesuit}$ energy flux 2--10\,keV calculated with spectrum by \textsc{cflux} in \textsc{Xspec}}\\
\multicolumn{9}{l}{$^{\clubsuit}$ photon flux 13.6\,eV--13.6\,keV calculated with nuclear primary spectrum (power-law with cut-off; not including the reflection part) by  }\\
\multicolumn{9}{l}{\quad \textsc{cpflux} in \textsc{Xspec}. This energy range is the same as that for $\xi_{\rm AD}$ \citep{2013ApJ...768..146G}.}\\
% photon flux at 1 keV (photons keV–1cm-2 s-1) of the cutoff broken power-law  only (no reflection) in the observed frame
\end{tabular}
\end{table*}

\subsubsection{Nuclear emission}\label{Gau2}
In this paper, we employed the state-of-the-art \textsc{relxill} model to model the reflection spectrum due to the X-ray illuminated relativistic accretion disk. This allows us to obtain statistical constraints on: a BH of spin 0.93$\pm  _{ 0.04 }^{0.04}$ with an AD inclination angle about $42^\circ$, emissivity indices varying from 4 to 9.6, ionisation parameter log($\xi_{\rm AD}$) varying in the range 2.7 to 3.5, a corona of photo-index varying between 1.73--2.25, and a high-energy cutoff $\sim 500$\,keV. The reflection fraction is 1.1--1.5 (implying the corona residing about 10--50$\, r_{\mathrm{g}}$ above the black hole\footnote{converted through the plot of ''reflection fraction v.s. height of corona'' in \cite{2016A&A...590A..76D}.}).

We compare our results with those in other studies: in the study of \cite{2015MNRAS.447.3029M} (\ngc by \nus), the AD inclination angle was reported as $<44^\circ$, consistent to what derived in this work; ionisation $\xi=22_{-7}^{+15}$, which is smaller than that for E11 in this work ($\log \xi \sim 2.7$); corona photo-index 1.9, which is consistent to what derived in this work; a very high-energy cutoff at $720^{+130}_{-190}$\,keV, which is higher than in this work, but still consistent with our results.

The radial index of the emissivity profile $\beta_{\rm AD}$ exhibits large errors. A test of making this parameter epoch-independent was performed. There was no great change in data-model residuals of each spectrum. The value of C-Statistic increased by 199, while the number of parameters decreased by 11.
C-Statistic/d.o.f: 12443/10154 = 1.225 (original); 12642/10165 = 1.244 (this test). A F-test was made to compare these two hypotheses. The restricted hypothesis (the index epoch-invariant) has a Chi-Squared of 11662 and a number of free parameters 99 (d.o.f 10165) while the unrestricted hypothesis (epoch-variant) has a Chi-Squared of 11498 and a number of free parameters 110 (d.o.f 10154).
We found that the hypothesis of epoch-dependent $\beta_{\rm AD}$ is better with a confidence level of more than 99\%. The best-fit values of $\beta_{\rm AD}$ are comprised between 1.4 and 3.1. The work of \cite{2010MNRAS.406.2013G} had an epoch-independent $\beta_{\rm AD}$ 1.9$\pm 0.3$.

\begin{figure}
\centering
\includegraphics[width=.32\textwidth,angle=270]{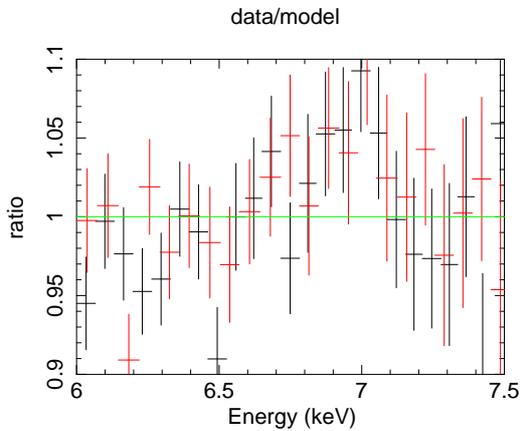}
\caption{Data/model ratio when the baseline model is applied to the XIS spectra of Epoch E08 in the 6--7.5 keV energy range.
}
\label{FGau6p7}
\end{figure}

Suzaku/XIS spectra show positive residuals at $\sim$7 keV (see an example of this feature in Fig.~\ref{FGau6p7}). This might be the signature of emission from resonant H-like iron lines. Therefore, we added a further, unresolved Gaussian profile to the model, constraining its rest-frame centroid energy to be comprised between 6.6 and 7\,keV. We apply the criterion explained in 
\cite{2002ApJ...571..545P}. The value of goodness-of-fit (C-Statistic) decreased by 190, while the number of free parameters increased by 3. C-Statistic/d.o.f: 12443/10154 = 1.225 (original); 12257/10151 = 1.207 (this test, adding a Gaussian profile). A F-test was made to compare these two hypotheses. The restricted hypothesis (without a Gaussian profile) has a Chi-Squared of 11498 and a number of free parameters 110 (d.o.f 10154) while the unrestricted hypothesis (with a Gaussian profile) has a Chi-Squared of 11336 and a number of free parameters 113 (d.o.f 10151).
We found that the model with this Gaussian profile is better at a confidence level of larger than 99\%.

\begin{figure}
\centering
\includegraphics[width=.30\textwidth,angle=270]{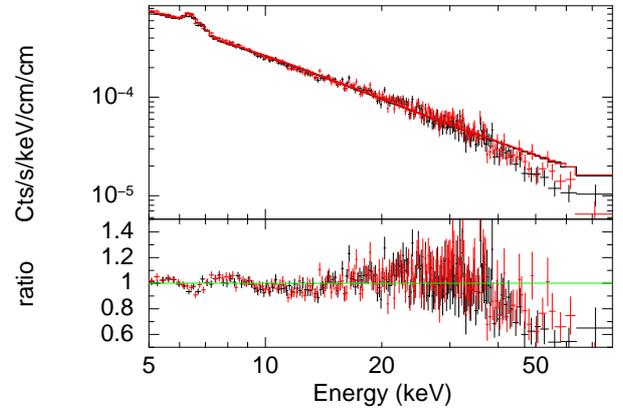}
\caption{The FPM spectra and the best-fit model without the relativistic disk reflection component.
}
\label{F0ReflFPM}
\end{figure}

\begin{figure}
\centering
\includegraphics[width=.30\textwidth,angle=270]{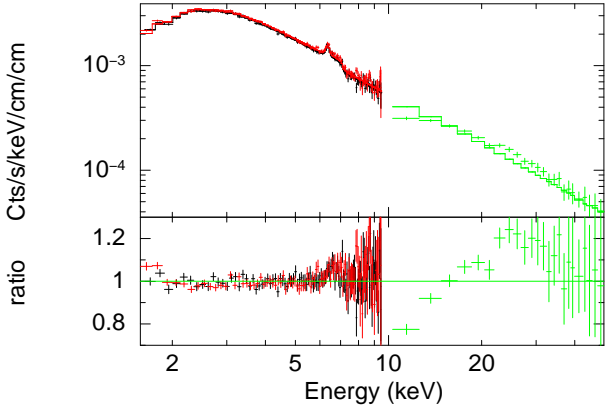}
\caption{The XIS and PIN spectra from Epoch 7, and the best-fit model without the relativistic disk reflection component.
}
\label{F0ReflXIS}
\end{figure}

In Figs.~\ref{F0ReflFPM} and \ref{F0ReflXIS}, we show the \nusc residuals once the normalisation of the \textsc{relxill} reflection component is set to zero, to show the effect that the relativistically smeared accretion disk reflection component has on the final best-fit model.

A Gaussian profile at about 1.7\,keV also is required by the fit. We do not have an explanation for this feature that we tentatively attribute to calibration uncertainties at the energies of the Si escape peak.

\subsubsection{Obscuration and torus}\label{}

The column density of the nuclear absorber varies in the range 2.8--3.8$\times$10$^{22}$~cm$^{-2}$.
\cite{2015MNRAS.447.3029M} reported $3.1^{+0.21}_{-0.20}\times$10$^{22}$~cm$^{-2}$, which agrees with the result in this work. \emph{Ginga} also observed variation in the column density in a timescale of $\sim$1--$\sim$10 days in 1991 \citep{bond1993soft}. This timescale is comparable with that presented in this paper, which corresponds to the coherent change timescale of an astrophysical structure of $\sim 0.001$--$\sim 0.01$\,pc. This size is similar to that of the broad-line region (BLR) of NGC\,5506, which can be estimated from the full width at half maximum of the Pa$\beta$ \citep{2002A&A...391L..21N}, $\sim 0.007$\,pc.

The best-fit inclination angle of the torus is about $43^\circ$. The matter in the torus does not intercept the line of sight, and reflects the nuclear emission. The column density of the torus is $\sim 1\times 10^{25}{\rm cm}^{-2}$. Note that this time-invariant column density is different from that on the LOS, $N_{\rm H}^{\rm nucl.}$.

Model \textsc{tbabs}[1] represents the medium with stable column density ($3.5 \times 10^{21}{\rm cm}^{-2}$) on the LOS from the outer soft-excess emitting region to the observer, including the medium causing the galactic extinction, inter-galactic medium, and galactic medium in NGC\,5506.

%\subsection{Correlation results}\label{}

%The fitting procedure employs C-Statistic \citep{1979ApJ...228..939C}. The goodness-of-fit is 12260 for 10158 degrees of freedom. 

%\subsection{Correlations with Eddington ratio}\label{}
\section{DISCUSSION AND CONCLUSIONS}\label{discuss}

\begin{figure}
\centering
\includegraphics[width=.45\textwidth]{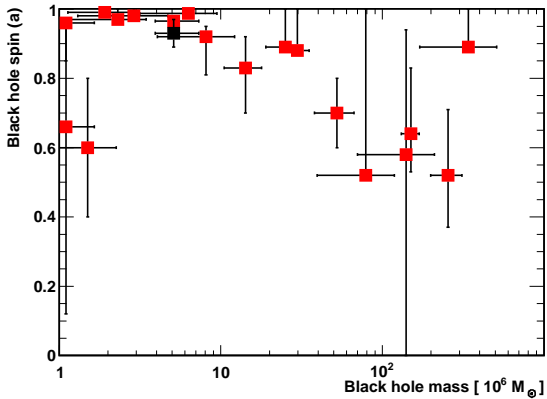}
\caption{Black hole mass/spin measurements of \ngc (in black; \citet{2009MNRAS.394.2141N}/this paper) and other AGNs (in red; data collected in \citet{2014SSRv..183..277R}.}                                                                                                                                                                           
\label{FMA}                                                                                                                                                                 
\end{figure}

In this paper, all the good quality X-ray archival data of \ngc are analyzed simultaneously to obtain an unprecedented combination of high signal-to-noise ratio and broadband energy coverage to achieve a good estimate of black hole spin. There have been several spin measurements for the SMBHs of AGNs through characterizing the relativistic reflection spectra in X-ray \citep{2014SSRv..183..277R}. For NGC\,5506, the SMBH spin had not been measured. In a previous study on \ngc \citep{2010MNRAS.406.2013G}, the data quality (on a smaller data set than discussed in this paper) did not allow to constrain the spin: varying the black hole spin from Schwarzschild to maximally rotating yielded a variation of the $\chi^2$ lower than 0.5. In this paper, we constrain the black hole spin 0.93$\pm  _{ 0.04 }^{0.04}$ at 90\% confidence level for one interesting parameter. The paper of \cite{2014SSRv..183..277R} reports several spin measurements, with uncertainties varying between 0.01 and 1; in this paper, the uncertainty 
range is 0.08. In Fig.~\ref{FMA}, the SMBH mass-spin distribution of \ngc with that of other AGNs collected in \cite{2014SSRv..183..277R} are compared. There is a trend, whereby smaller black hole masses correspond to higher black hole spin. While the statistics is still small, NGC\,5506 agrees with this trend. In the black hole spin measurements reported in Fig.~\ref{FMA}, we also note that there can be bias in the mass-spin distribution. A spinning black hole can increase radiative efficiency \citep[e.g ][]{2011ApJ...736..103B,2016MNRAS.458.2012V}. As a result, bright AGNs can have a higher probability to have rapid-spinning BHs \citep{2016AN....337..375F}, and AGNs with rapid-spinning BHs are easier to be detected, i.e. there can be over-population on the high-spin side in Fig.~\ref{FMA}.

In our model, six parameters ($^{\rm e.d.}P$'s) are epoch-dependent: $N_{\rm H}^{\rm nucl.}$, $\beta_{\rm AD}$, log$(\xi_{\rm AD})$, $\alpha_{\rm coro.}$, $F_{\rm refl.}$, and $F_{\rm prim.}^{\rm coro.}$. They are all variable over time scales probed by our data (see Tab.~\ref{Tevo2}). We estimate the minimum variability timescale for each parameter : find the minimal values among all the $(t_{i+1}-t_{i})\times ^{\rm e.d.}P_{\rm mean}/|^{\rm e.d.}P_{i+1}-^{\rm e.d.}P_{i}|$ ($t$ observation date, $i$ epoch index, $^{\rm e.d.}P_{\rm mean}$ the average of all the $^{\rm e.d.}P_{i}$'s. We find that the minimum variability timescales for $N_{\rm H}^{\rm nucl.}$, $\beta_{\rm AD}$, $\alpha_{\rm coro.}$, log$(\xi_{\rm AD})$, $F_{\rm refl.}$ and $F_{\rm prim.}^{\rm coro.}$ are about 110, 16, 30, 61, 14, and 5 days. Among these six parameters, $F_{\rm prim.}^{\rm coro.}$ has the shortest variability timescales.

\begin{table}
\small{
\caption{Constant-function fitting to epoch-dependent parameters.}\label{T:epoch}
\centering
\begin{tabular}{lll} 
\hline\hline
parameter & $\chi^2/n.d.f.$ & p-value \\                     
\hline
$N_{\rm H}^{\rm nucl.}$     &  (4.17$\times 10^{1}$)/11    &  8.45$\times 10^{-6}$\\ 
$\beta_{\rm AD}$            &  (8.34$\times 10^{2}$)/11    &  0.00\\   
$\alpha_{\rm coro.}$        &  (6.44$\times 10^{1}$)/11    &  5.33$\times 10^{-10}$\\ 
log$(\xi_{\rm AD})$         &  (3.33$\times 10^{2}$)/11    &  0.00\\ 
$F_{\rm refl.}$             &  (6.24$\times 10^{1}$)/11    &  1.27$\times 10^{-9}$\\               
$F_{\rm prim.}^{\rm coro.}$ &  (6.13$\times 10^{5}$)/11    &  0.00\\ 
\hline 
\multicolumn{3}{l}{number of entries: 11 (epochs)}\\
\multicolumn{3}{l}{number of parameter: 1 (constant-function)}\\
\multicolumn{3}{l}{See description for each parameter in Subsect.~\ref{model}.}\\
\end{tabular}                                           
}
\end{table}

The parameter reflection fraction varies between 1.1 and 1.5 with a constant-function fitting $p$-value close to null, which implies that the geometry of the primary source and reflection changes epoch by epoch. The corresponded heights of the source above the disk (in the simple lamp-post model) are about about 10--50$\, r_{\mathrm{g}}$. The height can change by a factor of 5.

% The correlation study on them shows that when the source is brighter (with higher energy flux), the corona spectrum is harder (smaller values of $\alpha_{\rm coro.}$), and the radial emissivity profile along the disk is shallower. A possible interpretation is that the AD reflection occurs primarily in the region where the emissivity profile is shallower and the material is less ionized (smaller values of $\beta_{\rm AD}$ and log$(\xi_{\rm AD})$), expected to be farther away from the black hole. One possibility to test this hypothesis is fitting the epoch-dependent spectra with a model where the radial emissivity profile is kept fixed to 3, and the innermost radius ($r_{\rm in}$) of the emitting region is left free for fitting. The resulted best-fit $r_{\rm in}$ has a Spearman's correlation coefficient -0.538 with log$(\xi_{\rm AD})$ (a confidence level of 87\%).

Spearman's ranking correlation coefficients between several pairs of parameters were calculated and reported in Tab.~\ref{T:Spearman}. It is found that $\alpha_{\rm coro.}$, log$(\xi_{\rm AD})$ and $N_{\rm H}^{\rm nucl.}$ are correlated with $F_{\rm prim.}^{\rm coro.}$; $\beta_{\rm AD}$ is anti-correlated with $F_{\rm prim.}^{\rm coro.}$; $F_{\rm refl.}$ is not correlated with $F_{\rm prim.}^{\rm coro.}$. The low values of Spearman $P$'s for $N_{\rm H}^{\rm nucl.}$ with the other parameters can result from intrinsic degeneracy between these parameters in the fit since the obscuration ($N_{\rm H}^{\rm nucl.}$) is far away from the nuclear region where the other parameters play roles. Among the other pairs of parameters, the correlation of $F_{\rm prim.}^{\rm coro.}$ versus log$(\xi_{\rm AD})$ is significant, which agrees with the expectation that the more photons illuminating an AD, the more the disk is ionized. Besides, $F_{\rm prim.}^{\rm coro.}$ also has a weak correlation with $\alpha_{\rm coro.}$ and a weak anti-correlation with $\beta_{\rm AD}$. 
This gives the following physical picture. A high flux state of corona occurs in higher probability with a soft spectral state, and in this condition the AD is more illuminated and more ionized. Hence a wider area of the innermost disk is fully ionized and does not contribute to the reflection, and the primary reflection region shifts outwards where the disk emissivity profile is shallower.
\begin{table}
%\small{
\caption{Spearman's ranking correlation coefficients between pairs of epoch-dependent parameters.}\label{T:Spearman}
\centering
\begin{tabular}{llcc} 
\hline\hline
                       &                         &          & two-tailed \\%& signi-\\
X                      & Y                       & $\rho$   & probability\\%& ficant$^{\clubsuit}$\\ 
                       &                         &          &  $P$-value \\%& or not\\
\hline
$F_{\rm prim.}^{\rm coro.}$&$N_{\rm H}^{\rm nucl.}$& 0.307 & 0.359\\
$F_{\rm prim.}^{\rm coro.}$&$\beta_{\rm AD}$       &-0.152 & 0.655\\
$F_{\rm prim.}^{\rm coro.}$&$\alpha_{\rm coro.}$   & 0.287 & 0.392\\
$F_{\rm prim.}^{\rm coro.}$&log$(\xi_{\rm AD})$    & 0.870 & 0.001\\
$F_{\rm prim.}^{\rm coro.}$&$F_{\rm refl.}$        & 0.225 & 0.506\\
$N_{\rm H}^{\rm nucl.}$    & $\beta_{\rm AD}$      &-0.364 & 0.270\\
$N_{\rm H}^{\rm nucl.}$    & $\alpha_{\rm coro.}$  & 0.552 & 0.077\\
$N_{\rm H}^{\rm nucl.}$    & log$(\xi_{\rm AD})$   & 0.483 & 0.131\\
$N_{\rm H}^{\rm nucl.}$    & $F_{\rm refl.}$       & 0.371 & 0.260\\
$\beta_{\rm AD}$           & $\alpha_{\rm coro.}$  &-0.406 & 0.214\\
$\beta_{\rm AD}$           & log$(\xi_{\rm AD})$   &-0.020 & 0.951\\
$\beta_{\rm AD}$           & $F_{\rm refl.}$       & 0.252 & 0.453\\
$\alpha_{\rm coro.}$       & log$(\xi_{\rm AD})$   & 0.319 & 0.337\\
$\alpha_{\rm coro.}$       & $F_{\rm refl.}$       &-0.026 & 0.938\\
log$(\xi_{\rm AD})$        & $F_{\rm refl.}$       & 0.122 & 0.719\\
\hline 
%\multicolumn{4}{l}{$^{\spadesuit}$ energy flux 2--10\,keV calculated by \textsc{cflux} in \textsc{Xspec}}\\
%\multicolumn{5}{l}{$^{\clubsuit}$ defined as "$P<0.32$" (a confidence level of more than 68\,$\%$)}\\
\end{tabular}                                           
%}
\end{table}

\begin{figure}
\centering
\includegraphics[width=.45\textwidth]{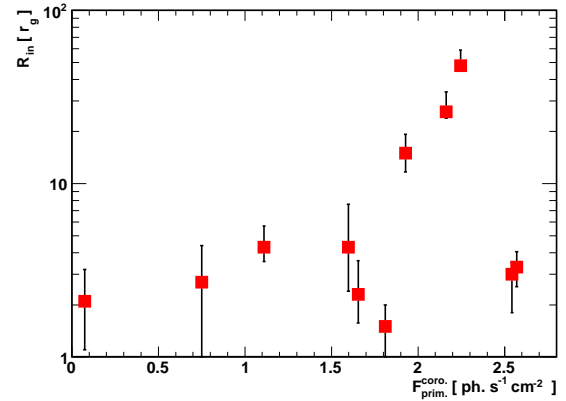}
\caption{Correlation between primary continuum flux and $r_{\rm in}$ in the \emph{modified model} (see the description in Sect.~\ref{discuss}).}                                                                                                                                                                           
\label{RF}                                                                                                                                                                 
\end{figure}

\begin{figure*}
  \centering
  \subfigure[]{
    \includegraphics[width=.32\textwidth]{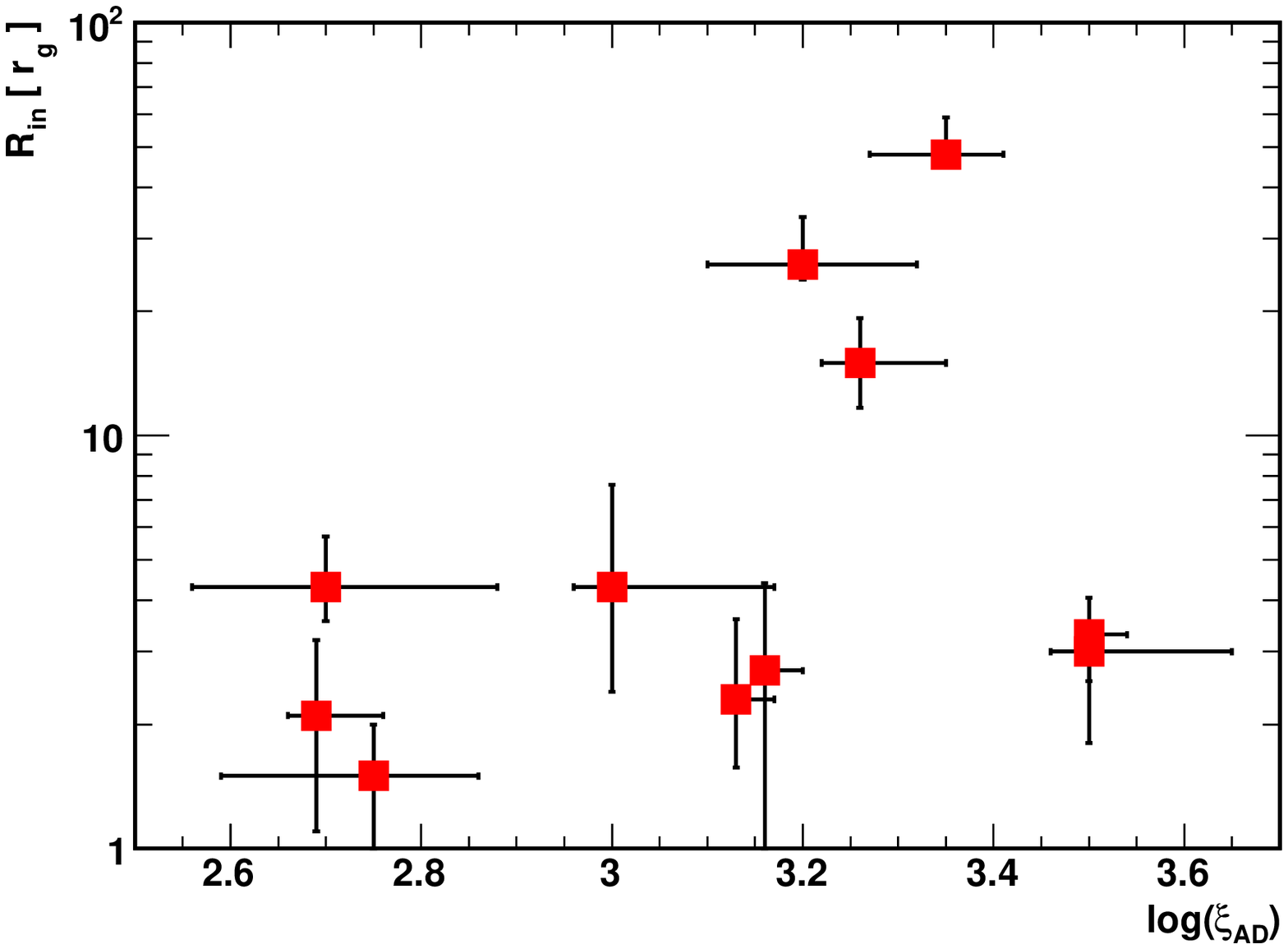}}     
  \subfigure[]{
    \includegraphics[width=.32\textwidth]{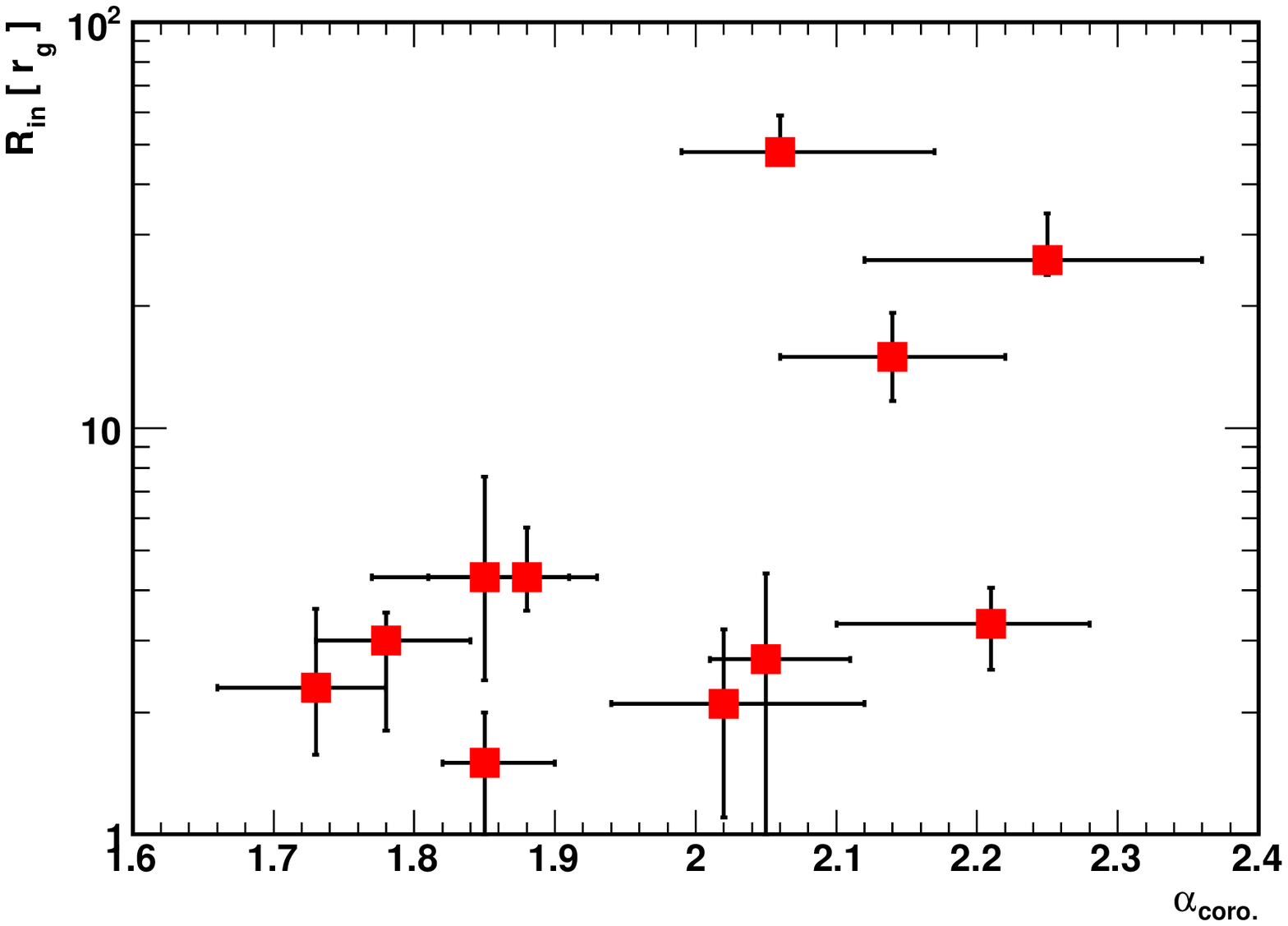}}
  \subfigure[]{
    \includegraphics[width=.32\textwidth]{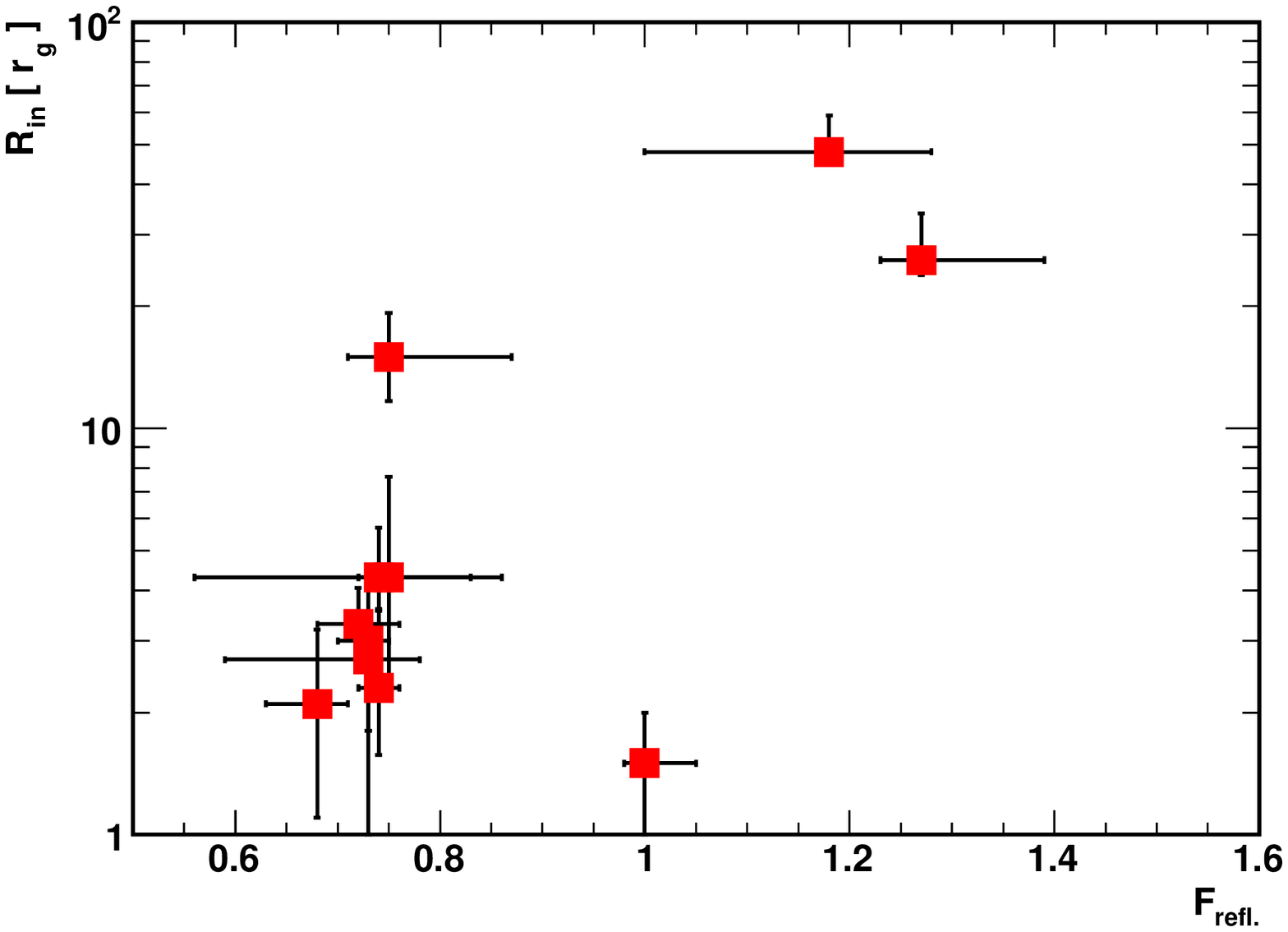}}      
  \caption{Relations between $r_{\rm in}$ in the \emph{modified model} (see the description in Sect.~\ref{discuss}) and other parameters.}
   \label{RFa}
\end{figure*}

Based on these correlations between parameters, there is a hypothesis to describe the astrophysical picture: what changes is the location of the innermost stable orbit due to over-ionisation of the innermost regions of the disk. At higher fluxes, a wider area of the innermost disk is fully ionized and does not contribute to the reflection when the illuminating source is closer to the disk. If this happens, the relative weight of disk regions increases that correspond to a lower ionisation level (i.e., ionisation parameter). Hence it came the idea of changing slightly the baseline model, assuming a constant index of the radial emissivity profile (3), and allowing the radius of the innermost stable orbit ($r_{\rm in}$) to vary at different epochs (\emph{modified model} hereafter). If this working hypothesis is correct, we should see a positive correlation between the primary continuum flux $F_{\rm prim.}^{\rm coro.}$ and $r_{\rm in}$ (See Fig.~\ref{RF}). The correlation between $r_{\rm in}$ and the flux has a Spearman correlation coefficient of 0.7 (a confidence level of 98\%). Formally there is a statistical linear correlation between the primary continuum flux and log($r_{\rm in}$). More than a linear correlation, there seems to be a special "high flux state" ($r_{\rm in}> 10$ $r_{\rm g}$) where the properties of the accretion flow differ from any other flux state. These three data points with $r_{\rm in}>$\,10\,$r_{\rm g}$ correspond to epochs 7, 8, and 9 when \suzc observed NGC\,5506. The relations ionisation log$(\xi_{\rm AD})$, continuum flux photo-index $\alpha_{\rm coro.}$, reflection fraction $F_{\rm refl.}$ vs $r_{\rm in}$ are shown in Fig.~\ref{RFa}. While the disk ionisation and $r_{\rm in}$ are, in general, positively correlated, at high ionisations (log$(\xi_{\rm AD})>3.2$) two (\xmnc observations, epochs 6 and 10) out of five data points are consistent to those measured at low ionisations. The reflection fraction $F_{\rm refl.}$ is expected to drop when the inner disc is over-
ionized, i.e. when the inner is larger. It is not seen in the results. A possible explanation is that the $F_{\rm refl.}$ related with not only the BH spin ($a$) and the size of the disk ($r_{\rm in}$ and $r_{\rm out}$), but also the height of the source (above the BH). In our case, the height does not positively correlate with $r_{\rm in}$, and the anti-correlation '$F_{\rm refl.}$--$r_{\rm in}$' does not hold.

We compare our results with another work \citep{2010MNRAS.406.2013G} on the reflection spectral characteristic of NGC\,5506, where spectra from the same \xmnc observations have been used, which has obtained the fit result $r_{\rm in}$ > 3$\, r_{\mathrm{g}}$ with an epoch-invariant emissivity profile of $\beta_{\rm AD}$=1.9. In order to compare with this result, We try a modified model (epoch-dependent $r_{\rm in}$) with a free epoch-invariant $\beta_{\rm AD}$. The fit results are reported in Table\,\ref{T:mod2}. Although the goodness of the fit is better than the original model, the best-fit $\beta_{\rm AD}$ is high ($>$3.5). Its best-fit black hole spin $a$ is consistent with that in the original model. To compare further with the results with $\beta_{\rm AD}=1.9$ in the work of \cite{2010MNRAS.406.2013G}, in our modified model where $r_{\rm in}$ is left free for fitting, we try the same emissivity profile, $\beta_{\rm AD}$=1.9, and check the BH spin $a$ in this case. We find a lower spin $a$ = 0.8. When we 
assume the canonical emissivity profile, $\beta_{\rm AD}$=3, an even lower spin $a$ = 0.68 is found. To explain why a smaller $\beta_{\rm AD}$ gives rise to a higher spin, it can be useful to compare the effect of the two parameters, $a$ and $\beta_{\rm AD}$ with the diagnostic diagrams 'Fe K-$\alpha$ line profiles vs different spin $a$ and different emissivity indices $\beta_{\rm AD}$': both a larger spin (closer to 1) and a larger $\beta_{\rm AD}$ (closer to 3) give rise to a broader Fe line. For this point, these two parameter can compensate with each other to some extent. This is why the modified model with $\beta_{\rm AD}$=3.0 results a smaller spin 0.68. These two models and the original model (with $a$=0.93) in our paper have the same number of degrees of freedom, so we can conveniently compare the fit statistics of them to judge which model is more favorable and which BH spin is more reasonable. The fit statistics is reported in Table\,\ref{T:mod2}. The original model with epoch-variant $\beta_{\rm AD}$, $a$ = 
0.93, and $r_{\rm in}$ = ISCO has the smallest fit statistics and fits the data the best, and is supposed to be our primary/final model. The motivation of the modified model is only to check whether the innermost disc may be over-ionized or not when the the primary irradiation is intensive. Therefore, our final result of spin measurement is based on the original model.

% \begin{table}
% \caption{Comparison of the original model and the modified models.}\label{T:mod2}
% \centering
% \begin{tabular}{lll} 
% \hline\hline
% model & $a$ & C-Statistic \\                     
% \hline
% original model $^{\spadesuit}$                        & 0.93$\pm  _{ 0.04 }^{0.04}$&  12184   \\ 
% modified model with $\beta_{\rm AD}$3.0 $^{\clubsuit}$& 0.68$\pm  _{ 0.25 }^{0.32}$&  12489   \\   
% modified model with $\beta_{\rm AD}$1.9 $^{\clubsuit}$& 0.80$\pm  _{ 0.06 }^{0.04}$&  12514   \\ 
% \hline 
% \multicolumn{3}{l}{all of them have the same degrees of freedom: 10154}\\
% \multicolumn{3}{l}{$^{\spadesuit}$ with epoch-variant $\beta_{\rm AD}$ and $r_{\rm in}$ = ISCO}\\
% \multicolumn{3}{l}{$^{\clubsuit}$ with epoch-variant $r_{\rm in}$}\\
% \end{tabular}       

\begin{table}
\caption{Comparison of the original model and the modified models.}\label{T:mod2}
\centering
\begin{tabular}{lccc} 
\hline\hline
model & $a$ & C-Statistic & $d.o.f.$\\                     
\hline
original model $^{\spadesuit}$                          & 0.93$\pm  _{ 0.04 }^{0.04}$&  12184  & 10154\\
modified model with a free $\beta_{\rm AD}$ $^{\sharp}$ & 0.90$\pm  _{ 0.03 }^{0.07}$&  12010  & 10153\\   
modified model with $\beta_{\rm AD}$3.0 $^{\clubsuit}$  & 0.68$\pm  _{ 0.25 }^{0.32}$&  12489  & 10154\\   
modified model with $\beta_{\rm AD}$1.9 $^{\clubsuit}$  & 0.80$\pm  _{ 0.06 }^{0.04}$&  12514  & 10154\\ 
\hline 
\multicolumn{4}{l}{all of them have the same degrees of freedom: 10154}\\
\multicolumn{4}{l}{$^{\spadesuit}$ with epoch-variant $\beta_{\rm AD}$ and $r_{\rm in}$ = ISCO}\\
\multicolumn{4}{l}{$^{\sharp}$ with epoch-variant $r_{\rm in}$ and epoch-invariant $\beta_{\rm AD}$; the best fit $\beta_{\rm AD}> 3.5$}\\
\multicolumn{4}{l}{$^{\clubsuit}$ with epoch-variant $r_{\rm in}$}\\
\end{tabular}   

\end{table}

Furthermore, The modified models, without the condition $r_{\rm in}$ = ISCO, constrain the spin parameter possibly less effectively than the original one, and the fitting of the spin parameter can be dominated by the epochs where the inner disc radius is smaller. In order to test it, we compute how much the fit statistics changes for each spectrum when the spin parameter changes. The original C-statistics is 12518. As a result, the epoch of \nusc observation (E11; with the lowest primary continuum flux) has the largest difference when parameter $a$ is changed from unity to null: $\Delta_{\rm E11}=5$ while the difference for all the epochs $\Delta_{\rm all}=7$. This shows that the spin measurement in the modified model indeed is primarily based on the epoch with the smallest inner disc radius. On the other hand, in the same test for the original model, each spectrum constrain the spin parameter more effectively. Therefore, it is more reasonable to determine the value of the spin parameter based on the 
original model.

% \begin{figure}
% \centering
% \includegraphics[width=.45\textwidth]{./RF3.eps}
% \caption{Correlation between energy flux and $r_{\rm in}$ in the \emph{modified model} (see the description in Sect.~\ref{discuss}).}                                                                                                                                                                           
% \label{RF}                                                                                                                                                                 
% \end{figure}

%In this modified model, the spin parameter is $a > 0.97$ at 90\% confidence level for one interesting parameter, which is consistent with $a=0.982\pm  _{ 0.001 }^{0.001}$ in the baseline model.

%In this modified model, our spectral modeling is inconsistent: the measurement of the black hole spin $a$ssumes that the disk reflection is produced down to the ISCO at all epochs. Therefore, we checked its impact on the determination of the spin: $a > 0.97$ (in the modified model) at 90\% confidence level for one interesting parameter, which is consistent with $a=0.982\pm  _{ 0.001 }^{0.001}$ in the baseline model. 
%We can see that the statistical accuracy in the measurement of the black hole spin $a$chieved by our study is already excellent, but the \emph{total} uncertainty is larger as reflected in modified model. This uncertainty (i.e., our degree of ignorance of the black hole spin in NGC\,5506) is dominated by our ignorance in the physics of the underlying accretion flow, which we have only superficially touched through the modified model. 
%Therefore, the main limitation to relativistic spectroscopy in this moment is due to the uncertainties in the physics, not to the signal-to-noise of the data. 

The best-fit nuclear column density varies between 2.8 and 3.3$\times$10$^{22}$~cm$^{-2}$, i.e. always in the Compton-thin regime. The obscuring matter covering the nucleus, however, has an averaged integrated column density about two orders of magnitude larger. Given the moderate inclination of the system, this suggests that we are looking at the system through a thin atmosphere grazing the surface of the torus., whose column density increases towards the equatorial plane.

Our best-fit inclination angles of AD or torus for NLSy1 \ngc are $40^{\circ} < \theta < 50^{\circ}$. However, \cite{2008MNRAS.386L..15D} assumed low inclination angles ($\lesssim 15^{\circ}$) and flat geometries of broad-line regions to explain the smaller FWHMs of broad lines in NLSy1. This assumption is different from our fit results. In the case of a lower inclination angle, the 'narrow' broad emission lines of \ngc can only be attributed to its smaller SMBH mass, which agrees with the study of \cite{2009MNRAS.394.2141N} supporting smaller $M_{\rm BH}$ of NGC\,5506 on the basis of its X-ray variability properties.

%($2\times 10^6 \Msun$)

\section*{Acknowledgments}
We thank the referee's useful comments for improving this article. This work was supported by the NSFC (Grant No.~U1531117 and 11305038), Fudan University (Grant No.~IDH1512060), and the Thousand Young Talents Program. S.S. also acknowledges support from the AHEAD program for his visit to INAF-Bologna in June 2016. C.B. also acknowledges support from the Alexander von Humboldt Foundation. In this work, the authors used the data supplied by NASA's High Energy Astrophysics Science Archive Research Center in the US. The authors also used the data products from the \xmn, \suz, \nus, and \cha, funded by the ESA, JAXA, NASA.

\bibliographystyle{mnras}

\bibliography{BibtexB}
% 
% \section*{Appendix}\label{}

\label{lastpage}

\end{document}